\documentclass[12pt]{amsart}

\copyrightinfo{2000}{American Mathematical Society}

\newcommand{\cA}{{\mathcal A}}
\newcommand{\cB}{{\mathcal B}}
\newcommand{\cD}{{\mathcal D}}
\newcommand{\cE}{{\mathcal E}}
\newcommand{\cL}{{\mathcal L}}
\newcommand{\cS}{{\mathcal S}}

\newcommand{\bN}{{\mathbb N}}
\newcommand{\bZ}{{\mathbb Z}}
\newcommand{\bQ}{{\mathbb Q}}
\newcommand{\bR}{{\mathbb R}}
\newcommand{\bC}{{\mathbb C}}

\numberwithin{equation}{section}

\newtheorem{Theorem}{Theorem}[section]
\newtheorem{Lemma}{Lemma}[section]
\newtheorem{Corollary}{Corollary}[section]
\newtheorem{Definition}{Definition}[section]

\newtheorem{Proposition}{Proposition}[section]

\author{A.~Yu.~Khrennikov}
\address{International Center for Mathematical Modelling in Physics
and Cognitive Sciences MSI, V\"axj\"o University, \ SE-351 95,
V\"axj\"o, \ Sweden. \ Phone: +046\,(470)\,708790 \, Fax: +046\,(470)\,84004}
\email{andrei.khrennikov@msi.vxu.se}

\author{V.~M.~Shelkovich}
\address{Department of Mathematics, St.-Petersburg State Architecture
and Civil Engineering University, \ 2 Krasnoarmeiskaya 4, 190005,
St. Petersburg, \ Russia. \ Phone: +7\,(812)\,2517549 \, Fax: +7\,(812)\,3165872}
\email{shelkv@vs1567.spb.edu}

\title[$p$-Adic wavelets and $p$-adic pseudo-differential operators]
{$p$-Adic multidimensional wavelets and their application to $p$-adic
pseudo-differential operators}

\thanks{This paper was supported in part by the grant
of The Swedish Royal Academy of Sciences on collaboration with
scientists of former Soviet Union.
The second author (V.~S.) was also supported in part by DFG Project 436 RUS 113/809
and Grant 05-01-04002-NNIOa of Russian Foundation for Basic Research.}



\date{ }

\begin{document}

\begin{abstract}
In this paper we study some problems related with the theory of
multidimensional $p$-adic wavelets in connection with the
theory of multidimensional $p$-adic pseudo-differential operators
(in the $p$-adic Lizorkin space). We introduce a new
class of $n$-dimensional $p$-adic compactly supported wavelets. In
one-dimensional case this class includes the Kozyrev $p$-adic
wavelets. These wavelets (and their Fourier transforms) form an
orthonormal complete basis in ${\cL}^2(\bQ_p^n)$. A criterion for a
multidimensional $p$-adic wavelet to be an eigenfunction for a
pseudo-differential operator is derived. We prove that these
wavelets are eigenfunctions of the Taibleson fractional operator.
Since many $p$-adic models use pseudo-differential operators
(fractional operator), these results can be intensively used in
applications. Moreover, $p$-adic wavelets are used to construct
solutions of linear and {\it semi-linear} pseudo-differential
equations.
\end{abstract}

\maketitle

\section{Introduction}
\label{s1}

There are a lot of papers where different applications of $p$-adic
analysis to physical problems, stochastics, cognitive sciences and
psychology are studied~\cite{Ar-Dr-V}--~\cite{Bik-V},
~\cite{Kh1}--~\cite{Koch3},~\cite{Vl-V-Z}--~\cite{V2} (see also the
references therein).

The field $\bQ_p$ of $p$-adic numbers is defined as
the completion of the field of rational numbers $\bQ$ with respect
to the non-Archimedean $p$-adic norm $|\cdot|_p$. This norm is
defined as follows: $|0|_p=0$; if an arbitrary rational number $x\ne 0$ is
represented as $x=p^{\gamma}\frac{m}{n}$, where $\gamma=\gamma(x)\in \bZ$,
and $m$ and $n$ are not divisible by $p$, then $|x|_p=p^{-\gamma}$.
This norm in $\bQ_p$ satisfies the strong triangle inequality
$|x+y|_p\le \max(|x|_p,|y|_p)$.

It is known that for the $p$-adic analysis related to the mapping
$\bQ_p \to \bC$, where $\bC$ is the field of complex numbers, the
operation of partial differentiation is {\it not defined\/}, and as
a result, large number of models connected with $p$-adic
differential equations use pseudo-differential operators and the
theory of $p$-adic distributions (generalized functions) (see the
above mentioned papers and books). In particular, fractional
operators $D^{\alpha}=f_{-\alpha}*$ are extensively used, where
$f_{\alpha}$ is the $p$-adic {\it Riesz kernel\/}, $*$ is a
convolution. However, in general, $D^{\alpha}\varphi \not\in
{\cD}(\bQ_p^n)$ for $\varphi\in {\cD}(\bQ_p^n)$, where
${\cD}(\bQ_p^n)$ is the space of test functions. Consequently, the
operation $D^{\alpha}f$ is well defined only for some distributions
$f\in {\cD}'(\bQ_p^n)$. For example, $D^{-1}$ is {\it defined\/}
only on the test functions such that
$\int_{\bQ_p}\varphi(x)\,dx=0$~\cite[IX.2]{Vl-V-Z}.

We recall that similar problems arise for the ``$\bC$-case''
fractional operators (where all functions and distributions are
complex or real valued defined on spaces with real or complex
coordinates): in general, the Schwartzian test function space
${\cS}(\bR^n)$ {\it is not invariant\/} under fractional
operators~\cite{Sam3},~\cite{Sam-Kil-Mar}. To solve this problem, in
the excellent papers of P.~I.~Lizorkin~\cite{Liz1},~\cite{Liz3} a
new type spaces {\it invariant\/} under fractional operators were
introduced (see also~\cite{Sam3},~\cite{Sam-Kil-Mar}).

Taking into account the problems mentioned above, in~\cite{Al-Kh-Sh3},
the $p$-adic Lizorkin spaces of test functions and distributions were
introduced, and in~\cite{Al-Kh-Sh3},~\cite{Al-Kh-Sh4} a class of
pseudo-differential operators (including the Taibleson fractional
operator) defined on them was constructed.
The Lizorkin spaces are {\it invariant\/} under our pseudo-differential
operators, and consequently, these spaces are their ``natural''
definition domains and can play a key role in considerations
related to the fractional operators problems.

Recall that for the one-dimensional case the orthonormal complete
basis of eigenfunctions (\ref{62.0}) of the Vladimirov operator
$D^{\alpha}$ was constructed by S.~V.~Kozyrev~\cite{Koz0}. The
eigenfunctions (\ref{62.0}) are $p$-adic compactly supported
wavelets. Further development and generalization of the theory of
such type wavelets can be found in the papers by
S.~V.~Kozyrev~\cite{Koz1}, ~\cite{Koz2}, A.~Yu.~Khrennikov, and
S.~V.~Kozyrev \cite{Kh-Koz1}, ~\cite{Kh-Koz2}, J.~J.~Benedetto, and
R.~L.~Benedetto~\cite{Ben-Ben}, and R.~L.~Benedetto \cite{Ben1}.

It is typical that such type $p$-adic compactly supported wavelets
are eigenfunctions of $p$-adic pseudo-differential operators.
Moreover, these wavelets satisfy the condition
$\int_{\bQ_p}\varphi(x)\,dx=0$ (see~\cite{Koz0}), and, in view of
Lemma~\ref{lem1}, belong to the Lizorkin space $\Phi(\bQ_p)$.
In~\cite{Al-Kh-Sh3}, there waqs derived the necessary and sufficient
condition for multidimensional $p$-adic pseudo-differential
operators to have such type multidimensional wavelets as
eigenfunctions. Thus the wavelets theory play a key role in $p$-adic
analysis.

{\bf Contents of the paper.} In this paper problems related with the
theory of multidimensional $p$-adic pseudo-differential operators
and the theory of multidimensional $p$-adic wavelets are studied.
Here the results of our paper~\cite{Al-Kh-Sh3} are intensively used.

In Sec.~\ref{s2}, we recall some facts from the $p$-adic theory of
distributions~\cite{G-Gr-P}, \cite{Taib1},
\cite{Taib3},~\cite{Vl-V-Z}. In Sec.~\ref{s3}, some facts from the
theory of the $p$-adic Lizorkin spaces~\cite{Al-Kh-Sh3} are
recalled. In Sec.~\ref{s4}, we recall some facts on the
multidimensional pseudo-differential operators defined in the
Lizorkin space of distributions ${\cD}'(\bQ_p^n)$. The fractional
Taibleson operator~\cite[\S2]{Taib1},~\cite[III.4.]{Taib3} is among
them. The Lizorkin spaces are {\it invariant\/} under our
pseudo-differential operators. It is appropriate to mention here
that the class of our operators includes the pseudo-differential
operators studied in~\cite{Koch3},~\cite{Z1},~\cite{Z2}.

In Sec.~\ref{s5}, a {\it new type of $p$-adic compactly supported
wavelets} (in one- dimensional (\ref{62.1}) and multidimensional
(\ref{62.8}) cases) are introduced. These wavelets belong to the
Lizorkin space of test functions. The Kozyrev one-dimensional
wavelets~\cite{Koz0} (see (\ref{62.0})) is a particular case of our
one-dimensional wavelets (\ref{62.1}). The {\it scaling function} of
wavelets (\ref{62.1}) is a characteristic function of the unit disc.
The {\it two-scale equation} (\ref{62.0-2}) for these wavelets
is presented. However, in this paper the multiresolution analysis is not
considered. The one-dimensional wavelets (\ref{62.1}) and
multidimensional wavelets (\ref{62.8}) form orthonormal complete
bases in ${\cL}^2(\bQ_p)$ and ${\cL}^2(\bQ_p^n)$, respectively (see
Theorems~\ref{th2},~\ref{th2.1}). Their Fourier transforms also form
orthonormal complete bases in ${\cL}^2(\bQ_p)$ and
${\cL}^2(\bQ_p^n)$, respectively (see Corollary~\ref{cor1},~\ref{cor2}).

In Sec.~\ref{s6}, the spectral theory of our pseudo-differential
operators is constructed. By Theorem~\ref{th4.1} the
criterion (\ref{64.1***}) for multidimensional $p$-adic
pseudo-differential operators (\ref{51}) to have multidimensional
wavelets (\ref{62.8}) as eigenfunctions is derived. In particular,
the multidimensional wavelets (\ref{62.8}) are eigenfunctions of the
Taibleson fractional operator (see (\ref{66})).

Since many $p$-adic models use pseudo-differential operators,
in particular, fractional operator, these results on $p$-adic wavelets can be
intensively used in applications. Moreover, $p$-adic wavelets can be used to
construct solutions of linear and {\it semi-linear} pseudo-differential
equations~\cite{Al-Kh-Sh5},~\cite{Kh-Koz3},~\cite{Koz-Os-Av-1}.

\section{$p$-Adic distributions}
\label{s2}

Now we recall some facts from the theory of $p$-adic distributions
(generalized functions).
We shall systematically use the notations and results from~\cite{Vl-V-Z}.
Let $\bN$, $\bZ$, $\bC$ be the sets of positive integers, integers,
complex numbers, respectively.
Denote by $\bQ_p^{*}=\bQ_p\setminus\{0\}$ the multiplicative group
of the field $\bQ_p$.
The space $\bQ_p^n=\bQ_p\times\cdots\times\bQ_p$ consists of points
$x=(x_1,\dots,x_n)$, where $x_j \in \bQ_p$, $j=1,2\dots,n$, \ $n\ge 2$.
The $p$-adic norm on $\bQ_p^n$ is
\begin{equation}
\label{8}
|x|_p=\max_{1 \le j \le n}|x_j|_p, \quad x\in \bQ_p^n.
\end{equation}

Denote by $B_{\gamma}^n(a)=\{x: |x-a|_p \le p^{\gamma}\}$ the ball
of radius $p^{\gamma}$ with the center at a point $a=(a_1,\dots,a_n)\in \bQ_p^n$
and by $S_{\gamma}^n(a)=\{x: |x-a|_p = p^{\gamma}\}
=B_{\gamma}^n(a)\setminus B_{\gamma-1}^n(a)$ its boundary (sphere),
$\gamma \in \bZ$. For $a=0$ we set $B_{\gamma}^n(0)=B_{\gamma}^n$ and
$S_{\gamma}^n(0)=S_{\gamma}^n$. For the case $n=1$ we will omit the
upper index $n$.
Here
\begin{equation}
\label{9}
B_{\gamma}^n(a)=B_{\gamma}(a_1)\times\cdots\times B_{\gamma}(a_n),
\end{equation}
where $B_{\gamma}(a_j)=\{x_j: |x_j-a_j|_p \le p^{\gamma}\}$ is a
disc of radius $p^{\gamma}$ with the center at a point $a_j\in
\bQ_p$, $j=1,2\dots,n$. Any two balls in $\bQ_p^n$ either are
disjoint or one contains the other. Every point of the ball is its
center.

According to~\cite[I.3,Examples 1,2.]{Vl-V-Z}, the disc $B_{\gamma}$
is represented by the sum of $p^{\gamma-\gamma'}$ {\it disjoint}
discs $B_{\gamma'}(a)$, $\gamma'<\gamma$:
\begin{equation}
\label{79.0}
B_{\gamma}=B_{\gamma'}\cup\cup_{a}B_{\gamma'}(a),
\end{equation}
where $a=0$ and
$a=a_{-r}p^{-r}+a_{-r+1}p^{-r+1}+\cdots+a_{-\gamma'-1}p^{-\gamma'-1}$
are the centers of the discs $B_{\gamma'}(a)$, \
$r=\gamma,\gamma-1,\gamma-2,\dots,\gamma'+1$, \, $0\le a_j\le p-1$,
\, $a_{-r}\ne 0$.

In particular, the disc $B_{0}$ is represented by the sum of $p$
{\it disjoint} discs
\begin{equation}
\label{79}
B_{0}=B_{-1}\cup\cup_{r=1}^{p-1}B_{-1}(r),
\end{equation}
where $B_{-1}(r)=\{x\in S_{0}: x_0=r\}=r+p\bZ_p$, $r=1,\dots,p-1$;
$B_{-1}=\{|x|_p\le p^{-1}\}=p\bZ_p$; and
$S_{0}=\{|x|_p=1\}=\cup_{r=1}^{p-1}B_{-1}(r)$. Here all the discs
are disjoint. We call covering (\ref{79.0}), (\ref{79}) the
{\it canonical covering} of the disc $B_{0}$.

A complex-valued function $f$ defined on $\bQ_p^n$ is called
{\it locally-constant} if for any $x\in \bQ_p^n$ there exists
an integer $l(x)\in \bZ$ such that
$$
f(x+y)=f(x), \quad y\in B_{l(x)}^n.
$$

Let ${\cE}(\bQ_p^n)$ and ${\cD}(\bQ_p^n)$ be the
linear spaces of locally-constant $\bC$-valued functions on $\bQ_p^n$
and locally-constant $\bC$-valued functions with compact supports
(so-called test functions), respectively; ${\cD}(\bQ_p)$,
${\cE}(\bQ_p)$~\cite[VI.1.,2.]{Vl-V-Z}.
If $\varphi \in {\cD}(\bQ_p^n)$, according to Lemma~1 from~\cite[VI.1.]{Vl-V-Z},
there exists $l\in \bZ$, such that
$$
\varphi(x+y)=\varphi(x), \quad y\in B_l^n, \quad x\in \bQ_p^n.
$$
The largest of such numbers $l=l(\varphi)$ is called the
{\it parameter of constancy} of the function $\varphi$.
Let us denote by ${\cD}^l_N(\bQ_p^n)$ the finite-dimensional space of
test functions from ${\cD}(\bQ_p^n)$ having supports in the ball $B_N^n$
and with parameters of constancy $\ge l$~\cite[VI.2.]{Vl-V-Z}.
Denote by ${\cD}'(\bQ_p^n)$ the set of all linear functionals on
${\cD}(\bQ_p^n)$~\cite[VI.3.]{Vl-V-Z}.

Let us introduce in ${\cD}(\bQ_p^n)$ a {\it canonical
$\delta$-sequence} $\delta_k(x)=p^{nk}\Omega(p^k|x|_p)$,
and a {\it canonical $1$-sequence}
$\Delta_k(x)=\Omega(p^{-k}|x|_p)$, $k \in \bZ$, \
$x\in \bQ_p^n$, where
\begin{equation}
\label{10}
\Omega(t)=\left\{
\begin{array}{lcr}
1, &&\quad 0 \le t \le 1, \\
0, &&\quad t>1. \\
\end{array}
\right.
\end{equation}
Here $\Delta_k(x)$ is the characteristic function of the ball $B_{k}^n$.
It is clear~\cite[VI.3., VII.1.]{Vl-V-Z} that
$\delta_k \to \delta$, $k \to \infty$ in ${\cD}'(\bQ_p^n)$
and $\Delta_k \to 1$, $k \to \infty$ in ${\cE}(\bQ_p^n)$.

The Fourier transform of $\varphi\in {\cD}(\bQ_p^n)$ is defined by the
formula
$$
F[\varphi](\xi)=\int_{\bQ_p^n}\chi_p(\xi\cdot x)\varphi(x)\,d^nx,
\quad \xi \in \bQ_p^n,
$$
where $\chi_p(\xi\cdot x)=e^{2\pi i\sum_{j=1}^{n}\{\xi_j x_j\}_p}$; \
$\xi\cdot x$ is the scalar product of vectors; the function
$\chi_p(\xi_j x_j)=e^{2\pi i\{\xi_j x_j\}_p}$ for every fixed
$\xi_j \in \bQ_p$ is an additive character of the field $\bQ_p$, \
$j=1,\dots,n$; $\{x\}_p$ is the {\it fractional part} of a number
$x\in \bQ_p$ which is defined as follows
\begin{equation}
\label{8.2**}
\{x\}_p=\left\{
\begin{array}{lll}
0,\quad \text{if} \quad \gamma(x)\ge 0 \quad  \text{or} \quad x=0,&&  \\
p^{\gamma}(x_0+x_1p+x_2p^2+\cdots+x_{|\gamma|-1}p^{|\gamma|-1}),
\quad \text{if} \quad \gamma(x)<0. && \\
\end{array}
\right.
\end{equation}

The Fourier transform is a linear isomorphism ${\cD}(\bQ_p^n)$ into
${\cD}(\bQ_p^n)$. Moreover, according
to~\cite[Lemma~A.]{Taib1},~\cite[III,(3.2)]{Taib3},
~\cite[VII.2.]{Vl-V-Z},
\begin{equation}
\label{12}
\varphi(x) \in {\cD}^l_N(\bQ_p^n) \quad \text{iff} \quad
F\big[\varphi(x)\big](\xi) \in {\cD}^{-N}_{-l}(\bQ_p^n).
\end{equation}

We define the Fourier transform $F[f]$ of a distribution
$f\in {\cD}'(\bQ_p^n)$ by the relation~\cite[VII.3.]{Vl-V-Z}:
\begin{equation}
\label{13}
\langle F[f],\varphi\rangle=\langle f,F[\varphi]\rangle,
\quad \forall \, \varphi\in {\cD}(\bQ_p^n).
\end{equation}

Let $A$ be a matrix and $b\in \bQ_p^n$. Then for a distribution
$f\in{\cD}'(\bQ_p^n)$ the following relation holds~\cite[VII,(3.3)]{Vl-V-Z}:
\begin{equation}
\label{14}
F[f(Ax+b)](\xi)
=|\det{A}|_p^{-1}\chi_p\big(-A^{-1}b\cdot \xi\big)F[f(x)]\big(A^{-1}\xi\big),
\end{equation}
where $\det{A} \ne 0$.
According to~\cite[IV,(3.1)]{Vl-V-Z},
\begin{equation}
\label{14.1}
F[\Delta_{k}](x)=\delta_{k}(x), \quad k\in \bZ, \qquad x \in \bQ_p^n.
\end{equation}
In particular, $F[\Omega(|\xi|_p)](x)=\Omega(|x|_p)$.

The convolution $f*g$ for distributions $f,g\in{\cD}'(\bQ_p^n)$ is
defined (see~\cite[VII.1.]{Vl-V-Z}) as
\begin{equation}
\label{11}
\langle f*g,\varphi\rangle
=\lim_{k\to \infty}\langle f(x)\times g(y),\Delta_k(x)\varphi(x+y)\rangle
\end{equation}
if the limit exists for all $\varphi\in {\cD}(\bQ_p^n)$, where
$f(x)\times g(y)$ is the direct product of distributions. If for
distributions $f,g\in {\cD}'(\bQ_p^n)$ the convolution $f*g$ exists
then~\cite[VII,(5.4)]{Vl-V-Z}
\begin{equation}
\label{15}
F[f*g]=F[f]F[g].
\end{equation}

It is well known (see, e.g.,~\cite[III.2.]{Vl-V-Z}) that any
{\it multiplicative character\/} $\pi$ of the field $\bQ_p$
can be represented as
$$
\pi(x)\stackrel{def}{=}\pi_{\alpha}(x)=|x|_p^{\alpha-1}\pi_{1}(x),
\quad x \in \bQ_p,
$$
where $\pi(p)=p^{1-\alpha}$ and $\pi_{1}(x)$ is a
{\it normed multiplicative character\/} such that
$\pi_1(x)=\pi_{1}(|x|_px)$, $\pi_1(p)=\pi_1(1)=1$, $|\pi_1(x)|=1$.
We denote $\pi_{0}=|x|_p^{-1}$.

\begin{Definition}
\label{de1.1} \rm
Let $\pi_{\alpha}$ be a multiplicative character of the field $\bQ_p$.

{\rm(a)} According to~\cite{Al-Kh-Sh1},~\cite{Al-Kh-Sh2}, a
distribution $f_m\in {\cD}'(\bQ_p)$ is said to be {\it associated
homogeneous {\rm(}in the wide sense{\rm)}\/} of
degree~$\pi_{\alpha}$ and order~$m$, \ $m \in \bN\cup\{0\}$, if
$$
\Bigl\langle f_m,\varphi\Big(\frac{x}{t}\Big)\Bigr\rangle
=\pi_{\alpha}(t)|t|_p \langle f_m,\varphi \rangle
+\sum_{j=1}^{m}\pi_{\alpha}(t)|t|_p\log_p^j|t|_p
\langle f_{m-j},\varphi \rangle
$$
for all $\varphi \in {\cD(\bQ_p)}$ and $t \in \bQ_p^*$, where
$f_{m-j}\in {\cD}'(\bQ_p)$ is an associated homogeneous distribution
of degree~$\pi_{\alpha}$ and order $m-j$, \ $j=1,2,\dots,m$, i.e.,
$$
f_m(tx)=\pi_{\alpha}(t)f_m(x)
+\sum_{j=1}^{m}\pi_{\alpha}(t)\log_p^j|t|_pf_{m-j}(x), \quad t \in \bQ_p^*.
$$
If $m=0$ we set that the above sum is empty.

{\rm(b)} We say that a distribution $f \in {\cD}'(\bQ_p^n)$ is
{\it associated homogeneous {\rm(}in the wide sense{\rm)}\/} of
degree $\pi_{\alpha}$ and order~$m$, \ $m \in \bN\cup\{0\}$, if for all
$t \in \bQ_p^*$ we have
$$
f_m(tx)=f_m(tx_1,\dots,tx_n)=\pi_{\alpha}(t)f_m(x)
+\sum_{j=1}^{m}\pi_{\alpha}(t)\log_p^j|t|_pf_{m-j}(x),
$$
where $\quad x=(x_1,\dots,x_n)\in \bQ_p^{n}$,  $f_{m-j}\in {\cD}'(\bQ_p^n)$
is an associated homogeneous distribution of degree~$\pi_{\alpha}$ and order
$m-j$, \ $j=1,2,\dots,m$.
An {\it associated homogeneous {\rm(}in the wide sense{\rm)}\/} distribution
of degree $\pi_{\alpha}(t)=|t|_p^{\alpha-1}$ and order~$m$ is called
{\it associated homogeneous} of degree~$\alpha-1$ and order~$m$.

{\rm(c)} An associated homogeneous distribution (in the wide sense)
of order $m=1$ is called {\it associated homogeneous} distribution
(see~\cite{Ge-Sh} and~\cite{Al-Kh-Sh1},~\cite{Al-Kh-Sh2}).

{\rm(d)} An associated homogeneous distribution of degree
$\pi_{\alpha}$ and order $m=0$ is called {\it homogeneous}
distribution of degree $\pi_{\alpha}$, i.e.,
$$
f_{0}(tx)=f_{0}(tx_1,\dots,tx_n)=\pi_{\alpha}(t)f_{0}(x),
\quad x=(x_1,\dots,x_n)\in \bQ_p^{n}.
$$
(for one-dimensional case see ~\cite[Ch.II,\S 2.3.]{G-Gr-P},
~\cite[VIII.1.]{Vl-V-Z}).
\end{Definition}

The multidimensional homogeneous distribution
$|x|_p^{\alpha-n}\in {\cD}'(\bQ_p^n)$
of degree $\alpha-n$ is constructed as follows.
If $Re\,\alpha>0$ then the function $|x|_p^{\alpha-n}$
generates a regular functional
\begin{equation}
\label{63.0}
\langle |x|_p^{\alpha-n},\varphi \rangle
=\int_{\bQ_p^n}|x|_p^{\alpha-n}\varphi(x)\,d^nx,
\quad \forall \, \varphi\in {\cD}(\bQ_p^n),
\end{equation}
where $|x|_p$, \ $x\in \bQ_p^n$ is given by (\ref{8}).
If  $Re\,\alpha \le 0$ this distribution is defined by means
of analytic continuation~\cite[(*)]{Taib1},~\cite[III,(4.3)]{Taib3},
~\cite[VIII,(4.2)]{Vl-V-Z}:
$$
\langle |x|_p^{\alpha-n},\varphi \rangle
=\int_{B_0^n}|x|_p^{\alpha-n}\big(\varphi(x)-\varphi(0)\big)\,d^nx
\qquad\qquad\qquad\qquad
$$
\begin{equation}
\label{63.1}
\qquad\qquad
+\int_{\bQ_p^n\setminus B_0^n}|x|_p^{\alpha-n}\varphi(x)\,d^nx
+\varphi(0)\frac{1-p^{-n}}{1-p^{-\alpha}},
\end{equation}
for all $\varphi\in {\cD}(\bQ_p^n)$, \ $\alpha\ne \mu_j=\frac{2\pi i}{\ln p}j$,
$j\in \bZ$.
The distribution $|x|_p^{\alpha-n}$ is an entire function of the complex
variable $\alpha$ everywhere except the points $\mu_j$, $j\in \bZ$,
where it has simple poles with residues $\frac{1-p^{-n}}{\log p}\delta(x)$.

Similarly to the one-dimensional case~\cite{Al-Kh-Sh1},~\cite{Al-Kh-Sh2},
one can construct the distribution $P(\frac{1}{|x|_p^{n}})$ called
the principal value of the function~$\frac{1}{|x|_p^{n}}$, $x\in \bQ_p^n$:
\begin{equation}
\label{63.1*}
\Bigl\langle P\Big(\frac{1}{|x|_p^{n}}\Big),\varphi \Bigr\rangle
=\int_{B_0^n}\frac{\varphi(x)-\varphi(0)}{|x|_p^{n}}\,d^nx
+\int_{\bQ_p^n\setminus B_0^n}\frac{\varphi(x)}{|x|_p^{n}}\,d^nx,
\end{equation}
for all $\varphi\in {\cD}(\bQ_p^n)$.
It is easy to show that this distribution is {\it associated
homogeneous\/} of degree $-n$ and order $1$
(see~\cite{Al-Kh-Sh1},~\cite{Al-Kh-Sh2}).

The Fourier transform of $|x|_p^{\alpha-n}$ is given by the
formula from~\cite[Theorem~2.]{Taib1},~\cite[III,Theorem~(4.5)]{Taib3},
~\cite[VIII,(4.3)]{Vl-V-Z}
\begin{equation}
\label{63.2}
F[|x|_p^{\alpha-n}]=\Gamma^{(n)}_p(\alpha)|\xi|_p^{-\alpha},
\quad \alpha \ne 0,\,n
\end{equation}
where the n-dimensional $\Gamma$-{\it function\/} $\Gamma^{(n)}_p(\alpha)$
is given by the following formulas (see~\cite[Theorem~1.]{Taib1},
~\cite[III,Theorem~(4.2)]{Taib3},~\cite[VIII,(4.4)]{Vl-V-Z}):
$$
\Gamma_p^{(n)}(\alpha)\stackrel{def}{=}
\lim_{k\to\infty}
\int_{p^{-k}\le |x|_p\le p^{k}} |x|_p^{\alpha-n}\chi_p(u\cdot x)\,d^nx
\qquad\quad
$$
\begin{equation}
\label{63.3}
=\int_{\bQ_p^n} |x|_p^{\alpha-n}\chi_p(x_1)\,d^nx
=\frac{1-p^{\alpha-n}}{1-p^{-\alpha}}
\end{equation}
where $|u|_p=1$, and the last integrals in the right-hand side of
(\ref{63.3}) are defined by means of analytic continuation with
respect to the parameter $\alpha$. Here
$\Gamma_p^{(1)}(\alpha)=\Gamma_p(\alpha)
=\int_{\bQ_p}|x|_p^{\alpha-1}\chi_p(x)\,dx
=\frac{1-p^{\alpha-1}}{1-p^{-\alpha}}$.

\section{The $p$-adic Lizorkin spaces}
\label{s3}

Let us introduce the $p$-adic {\it Lizorkin space of test functions\/}
$$
\Phi(\bQ_p^n)=\{\phi: \phi=F[\psi], \, \psi\in \Psi(\bQ_p^n)\},
$$
where
$$
\Psi(\bQ_p^n)
=\{\psi(\xi)\in \cD(\bQ_p^n): \psi(0)=0\}.
$$
Here $\Psi(\bQ_p^n), \Phi(\bQ_p^n)\subset \cD(\bQ_p^n)$.
The space $\Phi(\bQ_p^n)$ can be equipped with the topology
of the space $\cD(\bQ_p^n)$ which makes $\Phi(\bQ_p^n)$ a complete space.

In view of (\ref{12}), the following lemma holds.
\begin{Lemma}
\label{lem1}
{\rm (~\cite{Al-Kh-Sh3},~\cite{Al-Kh-Sh4})}
{\rm (a)} $\phi\in \Phi(\bQ_p^n)$ iff $\phi\in \cD(\bQ_p^n)$ and
\begin{equation}
\label{54}
\int_{\bQ_p^n}\phi(x)\,d^nx=0.
\end{equation}

{\rm (b)} $\phi \in {\cD}^l_N(\bQ_p^n)\cap\Phi(\bQ_p^n)$, i.e.,
$\int_{B^n_{N}}\phi(x)\,d^nx=0$,
iff \ $\psi=F^{-1}[\phi]\in {\cD}^{-N}_{-l}(\bQ_p^n)\cap\Psi(\bQ_p^n)$,
i.e., $\psi(\xi)=0$, $\xi \in B^n_{-N}$.
\end{Lemma}

In fact, for $n=1$, this lemma was proved in~\cite[IX.2.]{Vl-V-Z}.
Unlike the classical Lizorkin space,
any function $\psi(\xi)\in \Phi(\bQ_p^n)$ is equal to zero not only at $\xi=0$
but in a ball $B^n \ni 0$, as well.

Let $\Phi'(\bQ_p^n)$ denote the topological dual of the space
$\Phi(\bQ_p^n)$. We call it the $p$-adic {\it Lizorkin space of
distributions\/}.

By $\Psi^{\perp}$ and $\Phi^{\perp}$ we denote the
subspaces of functionals in $\cD'(\bQ_p^n)$ orthogonal to $\Psi(\bQ_p^n)$ and
$\Phi(\bQ_p^n)$, respectively. Thus
$\Psi^{\perp}=\{f\in \cD'(\bQ_p^n): f=C\delta, \, C\in \bC\}$ and
$\Phi^{\perp}=\{f\in \cD'(\bQ_p^n): f=C, \, C\in \bC\}$.

\begin{Proposition}
\label{pr2}
{\rm (~\cite{Al-Kh-Sh3})}
$$
\Phi'(\bQ_p^n)=\cD'(\bQ_p^n)/\Phi^{\perp}, \qquad
\Psi'(\bQ_p^n)=\cD'(\bQ_p^n)/\Psi^{\perp}.
$$
\end{Proposition}

The space $\Phi'(\bQ_p^n)$ can be obtained from $\cD'(\bQ_p^n)$ by
``sifting out'' constants. Thus two distributions in $\cD'(\bQ_p^n)$
differing by a constant are indistinguishable as elements of $\Phi'(\bQ_p^n)$.

Similarly to (\ref{13}), we define the Fourier transforms of
distributions $f\in \Phi_{\times}'(\bQ_p^n)$ and $g\in
\Psi_{\times}'(\bQ_p^n)$ by the relations:
\begin{equation}
\label{51}
\begin{array}{rcl}
\displaystyle
\langle F[f],\psi\rangle=\langle f,F[\psi]\rangle,
&& \forall \, \psi\in \Psi(\bQ_p^n), \medskip \\
\displaystyle
\langle F[g],\phi\rangle=\langle g,F[\phi]\rangle,
&& \forall \, \phi\in \Phi(\bQ_p^n). \\
\end{array}
\end{equation}
By definition, $F[\Phi(\bQ_p^n)]=\Psi(\bQ_p^n)$ and
$F[\Psi(\bQ_p^n)]=\Phi(\bQ_p^n)$, i.e., (\ref{51})
give well defined objects.

\section{Pseudo-differential operators in the Lizorkin space}
\label{s4}

\subsection{Pseudo-differential operators.}\label{s4.1}
Consider a class of pseudo-differential operators in the Lizorkin
space of the test functions $\Phi(\bQ_p^n)$
$$
(A\phi)(x)=F^{-1}\big[\cA(\xi)\,F[\phi](\xi)\big](x)
\qquad\qquad\qquad\qquad\qquad\qquad\qquad
$$
\begin{equation}
\label{64.3}
=\int_{\bQ_p^n}\int_{\bQ_p^n}\chi_p\big((y-x)\cdot \xi\big)
\cA(\xi)\phi(y)\,d^n\xi\,d^ny,
\quad \phi \in \Phi(\bQ_p^n)
\end{equation}
with symbols $\cA(\xi)\in \cE(\bQ_p^n\setminus \{0\})$.

\begin{Lemma}
\label{lem4}
The Lizorkin space $\Phi(\bQ_p^n)$ is invariant under the
pseudo- differential operators {\rm(\ref{64.3})}. Moreover,
$A(\Phi(\bQ_p^n))=\Phi(\bQ_p^n)$.
\end{Lemma}

\begin{proof}
In view of (\ref{12}) and results of Sec.~\ref{s3}, functions
$F[\phi](\xi)$ and $\cA(\xi)F[\phi](\xi)$ belong to $\Psi(\bQ_p^n)$,
and, consequently, $(A\phi)(x)\in \Phi(\bQ_p^n)$, i.e.,
$A(\Phi(\bQ_p^n))\subset \Phi(\bQ_p^n)$. Thus the
pseudo-differential operators (\ref{64.3}) are well defined, and the
Lizorkin space $\Phi(\bQ_p^n)$ is invariant under them. Moreover,
any function from $\Psi(\bQ_p^n)$ can be represented as
$\psi(\xi)=\cA(\xi)\psi_1(\xi)$, $\psi_1 \in \Psi(\bQ_p^n)$. This
implies that $A(\Phi(\bQ_p^n))=\Phi(\bQ_p^n)$.
\end{proof}

If we define a conjugate pseudo-differential operator $A^{T}$ as
\begin{equation}
\label{64.5}
(A^{T}\phi)(x)=F^{-1}[\cA(-\xi)F[\phi](\xi)](x)
=\int_{\bQ_p^n}\chi_p(-x\cdot \xi)\cA(-\xi)F[\phi](\xi)\,d^n\xi
\end{equation}
then one can define the operator $A$ in the Lizorkin space of
distributions: for $f \in \Phi'(\bQ_p^n)$ we have
\begin{equation}
\label{64.4}
\langle Af,\phi\rangle=\langle f,A^{T}\phi\rangle,
\qquad \forall \, \phi \in \Phi(\bQ_p^n).
\end{equation}
It is clear that
\begin{equation}
\label{64.3*}
Af=F^{-1}[\cA\,F[f]]\in \Phi'(\bQ_p^n),
\end{equation}
i.e., the Lizorkin space of distributions $\Phi'(\bQ_p^n)$
is invariant under pseudo-differential operators $A$.
Moreover, in view of Lemma~\ref{lem4}, $A(\Phi'(\bQ_p^n))=\Phi'(\bQ_p^n)$.

If $A, B$ are pseudo-differential operators with symbols
$\cA(\xi), \cB(\xi)\in \cE(\bQ_p^n\setminus \{0\})$, respectively,
then the operator $AB$ is well defined and represented by the formula
$$
(AB)f=F^{-1}[\cA\cB\,F[f]]\in \Phi'(\bQ_p^n).
$$
If $\cA(\xi)\ne 0$, $\xi\in \bQ_p^n\setminus \{0\}$ then we define
the inverse pseudo-differential operator by the formula
$$
A^{-1}f=F^{-1}[\cA^{-1}\,F[f]], \quad f\in \Phi'(\bQ_p^n).
$$

Thus the family of pseudo-differential operators $A$ with symbols
$\cA(\xi)\ne 0$, $\xi\in \bQ_p^n\setminus \{0\}$ forms an Abelian group.

If the symbol $\cA(\xi)$ of the operator $A$ is a {\it homogeneous\/} or
an {\it associated homogeneous\/} function (see Definition~\ref{de1.1})
then the pseudo-differential operator $A$ is called {\it homogeneous\/} or
{\it associated homogeneous\/}.

\subsection{The Taibleson fractional operator.}\label{s4.2}
Let us consider a pseudo- differential operator $D^{\alpha}_{x}$
with the symbol $\cA(\xi)=|\xi|_p^{\alpha}$. Thus, according to
(\ref{64.3}),
\begin{equation}
\label{61**}
\big(D^{\alpha}_{x}\phi\big)(x)
=F^{-1}\big[|\xi|^{\alpha}_pF[\phi](\xi)\big](x),
\quad \phi \in \Phi(\bQ_p^n).
\end{equation}
This multi-dimensional Taibleson fractional operator was introduced
in~\cite[\S2]{Taib1},~\cite[III.4.]{Taib3} on the space of distributions
${\cD}'(\bQ_p^n)$ for $\alpha\in \bC$, $\alpha\ne -n$.

In view of formulas (\ref{15}), (\ref{63.2}), (\ref{63.3}), the
relation (\ref{61**}) can be rewritten as a convolution
$$
\big(D^{\alpha}_{x}\phi\big)(x)\stackrel{def}{=}\kappa_{-\alpha}(x)*\phi(x)
=\langle \kappa_{-\alpha}(x),\phi(x-\xi)\rangle,
\quad x\in \bQ_p^n,
$$
where $\phi\in \Phi(\bQ_p^n)$, \ $\alpha \ne 0, \, -n$.
Here the distribution from ${\cD}'(\bQ_p^n)$
\begin{equation}
\label{63.4}
\kappa_{\alpha}(x)=\frac{|x|_p^{\alpha-n}}{\Gamma_p^{(n)}(\alpha)},
\quad \alpha \ne 0, \, \,  n, \qquad x\in \bQ_p^n,
\end{equation}
is called the multidimensional {\it Riesz
kernel\/}~\cite[\S2]{Taib1}, ~\cite[III.4.]{Taib3}, where the
function $|x|_p$, \ $x\in \bQ_p^n$ is given by (\ref{8}). The Riesz
kernel has a removable singularity at $\alpha=0$ and according
to~\cite[\S2]{Taib1},~\cite[III.4.]{Taib3},~\cite[VIII.2]{Vl-V-Z},
we obtain that $\langle
\kappa_{0}(x),\varphi(x)\rangle\stackrel{def}{=} \lim_{\alpha\to
0}\langle \kappa_{\alpha}(x),\varphi(x)\rangle=\varphi(0)$, for all
$\varphi\in {\cD}(\bQ_p^n)$, i.e.,
\begin{equation}
\label{63.5}
\kappa_{0}(x)\stackrel{def}{=}\lim_{\alpha\to 0}\kappa_{\alpha}(x)=\delta(x).
\end{equation}

Using (\ref{63.0}), (\ref{63.3}), (\ref{63.4}), and taking into
account (\ref{54}) (for details, see~\cite{Al-Kh-Sh3}, \cite{Al-Kh-Sh4}),
we define $\kappa_{n}(\cdot)$ as a distribution from the {\it Lizorkin
space of distributions\/} $\Phi'(\bQ_p^n)$:
\begin{equation}
\label{63.7}
\kappa_{n}(x)\stackrel{def}{=}\lim_{\alpha \to n}\kappa_{\alpha}(x)
=-\frac{1-p^{-n}}{\log p}\log|x|_p.
\end{equation}

With the help of (\ref{63.2}), (\ref{63.5}), (\ref{63.7}), it is easy
to see that
\begin{equation}
\label{63.9}
\kappa_{\alpha}(x)*\kappa_{\beta}(x)=\kappa_{\alpha+\beta}(x),
\quad \alpha, \beta \in \bC,
\end{equation}
holds in the sense of the Lizorkin space $\Phi'(\bQ_p^n)$.

In view of (\ref{63.5}), (\ref{63.7}), the multi-dimensional Taibleson
operator on the Lizorkin space of test functions is defined for all
$\alpha \in \bC$ as
\begin{equation}
\label{59**}
\big(D^{\alpha}_{x}\phi\big)(x)\stackrel{def}{=}\kappa_{-\alpha}(x)*\phi(x)
=\langle \kappa_{-\alpha}(x),\phi(x-\xi)\rangle,
\quad x\in \bQ_p^n,
\end{equation}
where $\phi\in \Phi(\bQ_p^n)$.

If $\alpha \ne n$ then the Riesz kernel $\kappa_{\alpha}(x)$ is a
{\it homogeneous\/} distribution of degree~$\alpha-n$, and if
$\alpha=n$ then the Riesz kernel is an {\it associated
homogeneous\/} distribution of degree $0$ and order $1$ (see
Definitions~\ref{de1.1},(b),(d)). Thus the Taibleson fractional
operator $D^{\alpha}_{x}$, $\alpha\ne -n$ is a {\it homogeneous\/}
pseudo-differential operator of degree~$\alpha$, and $D^{-n}_{x}$ is
an {\it associated homogeneous\/} pseudo-differential operator of
degree $-n$ and order $1$ with the symbol $\cA(\xi)=P(|\xi|_p^{-n})$
(see (\ref{63.1*})).

According to Lemma~\ref{lem4}, the Lizorkin space $\Phi(\bQ_p^n)$ is
invariant under the Taibleson fractional operator $D^{\alpha}_{x}$
and $D^{\alpha}_{x}(\Phi(\bQ_p^n))=\Phi(\bQ_p^n)$~\cite{Al-Kh-Sh3}.

In view of (\ref{64.5}), (\ref{64.4}), $(D^{\alpha}_{x})^{T}=D^{\alpha}_{x}$
and for $f\in \Phi'(\bQ_p^n)$ we have
\begin{equation}
\label{62**}
\langle D^{\alpha}_{x}f,\phi\rangle\stackrel{def}{=}
\langle f, D^{\alpha}_{x}\phi\rangle,
\quad \forall \, \phi\in \Phi(\bQ_p^n).
\end{equation}

It is clear that $D^{\alpha}_{x}(\Phi'(\bQ_p^n))=\Phi'(\bQ_p^n)$.
Moreover, in view of (\ref{63.9}), the family of operators
$D^{\alpha}_{x}$, $\alpha \in \bC$ on the Lizorkin space forms
an Abelian group: if $f \in \Phi'(\bQ_p^n)$ then
$D^{\alpha}_{x}D^{\beta}_{x}f=
D^{\beta}_{x}D^{\alpha}_{x}f=D^{\alpha+\beta}_{x}f$, \
$D^{\alpha}_{x}D^{-\alpha}_{x}f=f$, \ $\alpha,\beta \in \bC$.

\section{$p$-Adic wavelets}
\label{s5}

\subsection{One-dimensional $p$-adic wavelets.}\label{s5.1}
Let $n=1$. Consider the set
$$
I_p=\{a=p^{-\gamma}\big(a_{0}+a_{1}p+\cdots+a_{\gamma-1}p^{\gamma-1}\big):
\qquad\qquad\qquad\qquad
$$
\begin{equation}
\label{62.0**}
\qquad\qquad
\gamma\in \bN; \, a_j=0,1,\dots,p-1; \, j=0,1,\dots,\gamma-1\}.
\end{equation}
This set can be identified with the factor group $\bQ_p/\bZ_p$.
Let
$$
J_{p;m}=\{s=p^{-m}\big(s_{0}+s_{1}p+\cdots+s_{m-1}p^{m-1}\big):
\qquad\qquad\qquad\qquad
$$
\begin{equation}
\label{62.0*}
\qquad
s_j=0,1,\dots,p-1; \, j=0,1,\dots,m-1; s_0\ne 0\},
\end{equation}
where $m\ge 1$ is a {\it fixed} positive integer.

Let us introduce the function
$\theta_{s}^{(m)}(x)=\chi_p(sx)\Omega\big(|x|_p\big)$, $x\in \bQ_p$,
$s\in J_{p;m}$, and the functions generated by its dilatations
and translations:
\begin{equation}
\label{62.1}
\theta_{\gamma s a}^{(m)}(x)=p^{-\gamma/2}\chi_p\big(s(p^{\gamma}x-a)\big)
\Omega\big(|p^{\gamma}x-a|_p\big), \quad x\in \bQ_p,
\end{equation}
where $\gamma\in \bZ$, $s\in J_{p;m}$, $a\in I_p$,
$\Omega(t)$ is the characteristic function  (\ref{10}) of the
segment $[0,1]$.

Making the change of variables $\xi=p^{\gamma}x-a$
and taking into account (\ref{14.1}), we obtain
\begin{equation}
\label{62.1-1}
\int_{\bQ_p}\theta_{\gamma s a}^{(m)}(x)\,dx
=p^{\gamma/2}\int_{\bQ_p}\chi_p\big(s\xi\big)\Omega\big(|\xi|_p\big)\,d\xi
=p^{\gamma/2}\Omega\big(|s|_p\big)=0.
\end{equation}
Thus, in view of Theorem~\ref{th2} (see below), one can see that
the functions (\ref{62.1}) are {\it $p$-adic wavelets}.
Moreover, according to (\ref{62.1-1}) and Lemma~\ref{lem1}, the
functions $\theta_{\gamma s a}^{(m)}(x)$ belong to the Lizorkin space
$\Phi(\bQ_p)$.

It is clear that for any $\gamma\in \bZ$ and  $s\in J_{p;m}$
the functions (\ref{62.1-1}) are {\it periodical} with the periods
$T_{\gamma s}\in p^{m-\gamma}\bZ_p$.

In the case $m=1$, i.e., for $s=p^{-1}j$, $j=1,2,\dots,p-1$ these
wavelets coincide with the Kozyrev wavelets~\cite{Koz0}:
\begin{equation}
\label{62.0}
\theta_{\gamma s a}^{(1)}(x)=\theta_{\gamma j a}(x)
=p^{-\gamma/2}\chi_p\big(p^{-1}j(p^{\gamma}x-a)\big)
\Omega\big(|p^{\gamma}x-a|_p\big), \quad x\in \bQ_p,
\end{equation}
$\gamma\in \bZ$, $j=1,2,\dots,p-1$, $a\in I_p$.

In particular,
$\theta_{s}^{(1)}(x)=\theta_{j}(x)=\chi_p(p^{-1}jx)\Omega\big(|x|_p\big)$
for $j=1$. Since $|x|_p\le 1$, $x\in \bQ_p$, i.e.,
$x=x_{0}+x_{1}p+x_{2}p^2+\cdots$, we have
$p^{-1}x=p^{-1}x_{0}+x_{1}+x_{2}p+\cdots$, i.e., the fractional part
(\ref{8.2**}) of a number $p^{-1}x$ is equal to $\{p^{-1}x\}_p=p^{-1}x_{0}$.
According to (\ref{79}),
\begin{equation}
\label{62.0-1}
\theta_{1}(x)=\chi_p(p^{-1}x)\Omega\big(|x|_p\big)
=\left\{
\begin{array}{lll}
0, && |x|_p\ge p, \\
e^{2\pi i\frac{r}{p}}, && x\in B_{-1}(r), \, r=1,\dots,p-1, \\
1, && x\in B_{-1}. \\
\end{array}
\right.
\end{equation}
Thus the function
$\theta_{1}(x)=\chi_p(p^{-1}x)\Omega\big(|x|_p\big)$ takes values in
the set $\{0,e^{2\pi i\frac{r}{p}}:r=0,1,\dots,p-1\}$ of $p+1$ elements.

Now we consider
$\theta_{s}^{(1)}(x)=\theta_{j}(x)=\chi_p(p^{-1}jx)\Omega\big(|x|_p\big)$.
Since $|jx|_p\le 1$, $x\in \bQ_p$, we have
$jx=y_{0}+y_{1}p+y_{2}p^2+\cdots$,
$p^{-1}jx=p^{-1}y_{0}+y_{1}+y_{2}p+\cdots$, and
$\{p^{-1}jx\}_p=p^{-1}y_{0}$ (see (\ref{8.2**})).
Thus,
$$
\theta_{j}(x)=\chi_p(p^{-1}jx)\Omega\big(|x|_p\big)
=\left\{
\begin{array}{lll}
0, && |x|_p\ge p, \\
e^{2\pi i\{\frac{jr}{p}\}_p}, && x\in B_{-1}(r), \, r=1,\dots,p-1, \\
1, && x\in B_{-1}. \\
\end{array}
\right.
$$

It is clear that for the Kozyrev wavelets the {\it scaling function}
is the characteristic function of the unit disc $\Delta_0(x)=\Omega\big(|x|_p\big)$,
$x\in \bQ_p$, and in view of (\ref{79}) it satisfies the {\it two-scale equation}:
\begin{equation}
\label{62.0-2}
\Delta_0(x)=p^{-1/2}\sum_{r=0}^{p-1}h_{r}\Delta_0\Big(\frac{1}{p}x-\frac{r}{p}\Big),
\quad x\in \bQ_p,
\end{equation}
where $h_{r}=p^{1/2}$.
Relations (\ref{62.0-1}), (\ref{62.0-2}) imply that
\begin{equation}
\label{62.0-3}
\theta_{1}(x)=\chi_p(p^{-1}x)\Omega\big(|x|_p\big)
=p^{-1/2}\sum_{r=0}^{p-1}{\tilde h}_r\Delta_0\Big(\frac{1}{p}x-\frac{r}{p}\Big),
\quad x\in \bQ_p,
\end{equation}
where ${\tilde h}_r=p^{1/2}e^{2\pi i\frac{r}{p}}$, $r=0,1,\dots,p-1$.
Similarly to (\ref{62.0-3}), we have
\begin{equation}
\label{62.0-4}
\theta_{j}(x)=\chi_p(p^{-1}jx)\Omega\big(|x|_p\big)
=p^{-1/2}\sum_{r=0}^{p-1}{\tilde h}_r\Delta_0\Big(\frac{1}{p}x-\frac{r}{p}\Big),
\quad x\in \bQ_p,
\end{equation}
where ${\tilde h}_r=p^{1/2}e^{2\pi i\{\frac{jr}{p}\}_p}$, $r=0,1,\dots,p-1$.

In the same way we consider the function
$\theta_{s}^{(m)}(x)=\chi_p(sx)\Omega\big(|x|_p\big)$, $s\in J_{p;m}$.
Let $B_{0}=\cup_{a}B_{-m}(a)\cup B_{-m}$ be the {\it canonical covering}
(\ref{79.0}) of the disc $B_{0}$ with $p^{m}$ discs, $m\ge 1$, where
$a=0$ and $a=a_{r}p^{r}+a_{r+1}p^{r+1}+\cdots+a_{m-1}p^{m-1}$
is the center of the discs $B_{-m}$ and $B_{-m}(a)$, respectively,
\ $r=0,1,2,\dots,m-1$, \, $0\le a_j\le p-1$, \, $a_r\ne 0$.

For $x\in B_{-m}(a)$, $s\in J_{p;m}$, we have
$x=a+p^{m}\big(y_{0}+y_{1}p+y_{2}p^2+\cdots\big)$,
$s=p^{-m}\big(s_{0}+s_{1}p+\cdots+s_{m-1}p^{m-1}\big)$, $s_0\ne 0$;
$sx=sa+\xi$, $\xi\in \bZ_p$;
and $\{sx\}_p=\{sa\}_p=\{p^{r-m}
\big(a_{r}+a_{r+1}p+\cdots+a_{m-1}p^{m-r-1}\big)
\big(s_{0}+s_{1}p+\cdots+s_{m-1}p^{m-1}\big)\}_p$, \
$r=0,1,2,\dots,m-1$, (see (\ref{8.2**})).
Thus,
$$
\theta_{s}^{(m)}(x)=\chi_p(sx)\Omega\big(|x|_p\big)
\qquad\qquad\qquad\qquad\qquad\qquad\qquad\qquad
$$
$$
\qquad
=\left\{
\begin{array}{lll}
0, && |x|_p\ge p, \\
e^{2\pi i\{sa\}_p}, && x\in B_{-m}(a), \quad
a=\sum_{l=r}^{m-1}a_{l}p^{l}, \\
1, && x\in B_{-m}, \\
\end{array}
\right.
$$
where $s=p^{-m}\big(s_{0}+s_{1}p+\cdots+s_{m-1}p^{m-1}\big)$,
$0\le s_j\le p-1$, $j=0,1,\dots,m-1$, $s_0\ne 0$; \,
$a=a_{r}p^{r}+a_{r+1}p^{r+1}+\cdots+a_{m-1}p^{m-1}$,
$0\le a_j\le p-1$, $a_r\ne 0$, $r=0,1,\dots,m-1$.
Thus the function $\theta_{s}^{(m)}(x)=\chi_p(sx)\Omega\big(|x|_p\big)$
takes values in the set $\{0,1,e^{2\pi i\{sa\}_p}\}$ of $p^m+1$ elements.

In this case, using the {\it scaling function}, we obtain
\begin{equation}
\label{62.0-5}
\theta_{s}^{(m)}(x)=\chi_p(sx)\Omega\big(|x|_p\big)
=p^{-m/2}\sum_{a}{\tilde h}_a\Delta_0\Big(\frac{1}{p^m}x-\frac{a}{p^m}\Big),
\end{equation}
$x\in \bQ_p$, where ${\tilde h}_0=p^{m/2}$; \, ${\tilde h}_a=p^{m/2}e^{2\pi i\{sa\}_p}$, \,
$a=a_{r}p^{r}+a_{r+1}p^{r+1}+\cdots+a_{m-1}p^{m-1}$,
$r=0,1,\dots,m-1$, $0\le a_j\le p-1$, $a_r\ne 0$.

\begin{Theorem}
\label{th2}
The functions {\rm(\ref{62.1})} form an orthonormal complete
basis in ${\cL}^2(\bQ_p)$ {\rm(}$p$-adic wavelet basis{\rm)}.
\end{Theorem}

\begin{proof}
Consider the scalar product
$$
\big(\theta_{\gamma' s' a'}^{(m)}(x),\theta_{\gamma s a}^{(m)}(x)\big)
=p^{-(\gamma+\gamma')/2}
\qquad\qquad\qquad\qquad\qquad\qquad\qquad\qquad
$$
\begin{equation}
\label{62.3}
\times
\int_{\bQ_p}
\chi_p\big(s'(p^{\gamma'}x-a')-s(p^{\gamma}x-a)\big)
\Omega\big(|p^{\gamma}x-a|_p\big)\Omega\big(|p^{\gamma'}x-a'|_p\big)\,dx.
\end{equation}
If $\gamma\le\gamma'$, according to formula~\cite[VII.1]{Vl-V-Z},~\cite{Koz0}
\begin{equation}
\label{62.4-1}
\Omega\big(|p^{\gamma}x-a|_p\big)\Omega\big(|p^{\gamma'}x-a'|_p\big)
=\Omega\big(|p^{\gamma}x-a|_p\big)\Omega\big(|p^{\gamma'-\gamma}a-a'|_p\big),
\end{equation}
(\ref{62.3}) can be rewritten as
$$
\big(\theta_{\gamma' s' a'}^{(m)}(x),\theta_{\gamma s a}^{(m)}(x)\big)
=p^{-(\gamma+\gamma')/2}\Omega\big(|p^{\gamma'-\gamma}a-a'|_p\big)
\qquad\qquad\qquad\qquad\qquad
$$
\begin{equation}
\label{62.4}
\times
\int_{\bQ_p}
\chi_p\big(s'(p^{\gamma'}x-a')-s(p^{\gamma}x-a)\big)
\Omega\big(|p^{\gamma}x-a|_p\big)\,dx.
\end{equation}

Let $\gamma<\gamma'$. Making the change of variables $\xi=p^{\gamma}x-a$
and taking into account (\ref{14.1}), from (\ref{62.4}) we obtain
$$
\big(\theta_{\gamma' s' a'}^{(m)}(x),\theta_{\gamma s a}^{(m)}(x)\big)
=p^{-(\gamma+\gamma')/2}
\chi_p\big(s'(p^{\gamma'-\gamma}a-a')\big)
\qquad\qquad\qquad\qquad\quad
$$
$$
\qquad\qquad
\times
\Omega\big(|p^{\gamma'-\gamma}a-a'|_p\big)
\int_{\bQ_p}\chi_p\big((p^{\gamma'-\gamma}s'-s)\xi\big)
\Omega\big(|\xi|_p\big)\,d\xi
$$
\begin{equation}
\label{62.5}
=p^{-(\gamma+\gamma')/2}
\chi_p\big(s'(p^{\gamma'-\gamma}a-a')\big)
\Omega\big(|p^{\gamma'-\gamma}a-a'|_p\big)
\Omega\big(|p^{\gamma'-\gamma}s'-s|_p\big).
\end{equation}
Since
$$
\begin{array}{rclrcl}
\displaystyle
p^{\gamma'-\gamma}s'&=&p^{\gamma'-\gamma-m}\big(s_{0}'
+s_{1}'p+\cdots+s_{\gamma-1}'p^{m-1}\big), \smallskip \\
s&=&p^{-m}\big(s_{0}+s_{1}p+\cdots+s_{\gamma-1}p^{m-1}\big), \\
\end{array}
$$
where $s_{0}',s_{0}\ne 0$, $\gamma'-\gamma\le 1$, it is clear that
fractional part $\{p^{\gamma'-\gamma}s'-s\}_p\ne 0$. Thus
$\Omega\big(|p^{\gamma'-\gamma}s'-s|_p\big)=0$ and
$\big(\theta_{\gamma' s' a'}^{(m)}(x),\theta_{\gamma s a}^{(m)}(x)\big)=0$.

Consequently, the scalar product $\big(\theta_{\gamma' s'
a'}^{(m)}(x),\theta_{\gamma s a}^{(m)}(x)\big)=0$ can be nonzero
only if $\gamma=\gamma'$. In this case (\ref{62.5}) implies
\begin{equation}
\label{62.6}
\big(\theta_{\gamma s' a'}^{(m)}(x),\theta_{\gamma s a}^{(m)}(x)\big)
=p^{-\gamma}
\chi_p\big(s'(a-a')\big)\Omega\big(|a-a'|_p\big)\Omega\big(|s'-s|_p\big),
\end{equation}
where $\Omega\big(|a-a'|_p\big)=\delta_{a'a}$,
$\Omega\big(|s'-s|_p\big)=\delta_{s's}$, and $\delta_{s's}$,
$\delta_{a'a}$ are the Kronecker symbols.

Since $\int_{\bQ_p}\Omega\big(|p^{\gamma}x-a|_p\big)\,dx=p^{\gamma}$
~\cite[IV,(2.3)]{Vl-V-Z}, formulas (\ref{62.5}), (\ref{62.6}) imply that
\begin{equation}
\label{62.7}
\big(\theta_{\gamma' s' a'}^{(m)}(x),\theta_{\gamma s a}^{(m)}(x)\big)
=\delta_{\gamma'\gamma}\delta_{s's}\delta_{a'a}.
\end{equation}
Thus the system of functions (\ref{62.1}) is orthonormal.

To prove the completeness of the system of functions (\ref{62.1}),
we repeat the corresponding proof~\cite{Koz0} almost word for word.
Recall that the system of the characteristic functions of the discs
$B_{k}$ is complete in ${\cL}^2(\bQ_p)$. Consequently, taking into
account that the system of functions $\{\theta_{\gamma s
a}^{(m)}(x):\gamma\in \bZ,s\in J_{p;m},a\in I_p\}$ is invariant
under dilatations and translations, in order to prove that it is a
complete system, it is sufficient to verify the Parseval identity
for the characteristic function $\Omega(|x|_p)$.

If $0\le \gamma$, according to (\ref{62.4-1}), (\ref{14.1}),
$$
\big(\Omega(|x|_p),\theta_{\gamma s a}^{(m)}(x)\big)
=p^{-\gamma/2}\Omega\big(|-a|_p\big)
\int_{\bQ_p}\chi_p\big(s(p^{\gamma}x-a)\big)\Omega\big(|x|_p\big)\,dx
\qquad\qquad
$$
$$
\qquad
=p^{-\gamma/2}\chi_p\big(-sp^{\gamma}a)\big)
\Omega\big(|sp^{\gamma}|_p\big)\Omega\big(|-a|_p\big)
=\left\{
\begin{array}{lll}
0, && a\ne 0, \\
0, && a=0, \, \gamma \le m-1, \\
p^{-\gamma/2}, && a=0, \, \gamma \ge m. \\
\end{array}
\right.
$$

If $0> \gamma$, according to (\ref{62.4-1}), (\ref{14.1}),
$$
\big(\Omega(|x|_p),\theta_{\gamma s a}^{(m)}(x)\big)
=p^{-\gamma/2}\Omega\big(|p^{-\gamma}a|_p\big)
\int_{\bQ_p}\chi_p\big(s(p^{\gamma}x-a)\big)
\Omega\big(|p^{\gamma}x-a|_p\big)\,dx
$$
$$
=p^{-\gamma/2}\Omega\big(|p^{-\gamma}a|_p\big)
\int_{\bQ_p}\chi_p\big(s\xi\big)\Omega\big(|\xi|_p\big)\,d\xi
=p^{-\gamma/2}\Omega\big(|p^{-\gamma}a|_p\big)\Omega\big(|s|_p\big)=0.
$$
Thus,
$$
\sum_{\gamma\in \bZ,s\in J_{p;m},a\in I_p}
\big|\big(\Omega(|x|_p),\theta_{\gamma s a}^{(m)}(x)\big)\big|^2
=\sum_{\gamma=m}^{\infty}\sum_{s\in J_{p;m}}p^{-\gamma}
\qquad\qquad\qquad\qquad\qquad
$$
$$
\qquad\qquad
=p^{m-1}(p-1)\frac{p^{-m}}{1-p^{-1}}=1
=\big|\big(\Omega(|x|_p),\Omega(|x|_p)\big|^2.
$$

Thus the system of functions (\ref{62.1}) is an orthonormal basis in
${\cL}^2(\bQ_p)$ ($p$-adic wavelet basis).
\end{proof}

\begin{Corollary}
\label{cor1}
The functions
$$
{\widetilde\theta}_{\gamma s a}^{(m)}
=F[\theta_{\gamma s a}^{(m)}](\xi)
=p^{\gamma/2}\chi_p\big(p^{-\gamma}a\cdot\xi\big)
\Omega\big(|s+p^{-\gamma}\xi|_p\big),\quad \xi\in \bQ_p,
$$
form an orthonormal complete basis in ${\cL}^2(\bQ_p)$, $a\in I_p$;
$s\in J_{p;m}$; $m\ge 1$ is a {\it fixed} positive integer.
\end{Corollary}

The proof follows from Theorem~\ref{th2}, formula (\ref{64.8*})
(see below) and the Parseval formula~\cite[VII,(4.1)]{Vl-V-Z}

\subsection{Multidimensional $p$-adic wavelets.}\label{s5.2}
Let us introduce $n$-dimensional functions generated by the $n$-direct
product of the one-dimensional $p$-adic wavelets (\ref{62.1}):
\begin{equation}
\label{62.8}
\Theta_{\gamma s a}^{(m)}(x)=p^{-n\gamma/2}
\chi_p\big(s\cdot(p^{\gamma}x-a)\big)
\Omega\big(|p^{\gamma}x-a|_p\big),
\end{equation}
$x=(x_1,\dots,x_n)\in \bQ_p^n$, where $\gamma\in \bZ$;
$a=(a_1,\dots,a_n)\in I_p^n$; $s=(s_1,\dots,s_n)\in J_{p;m}^n$;
$m=(m_1,\dots,m_n)$, $m_j\ge 1$ is a {\it fixed} positive
integer, $j=1,2,\dots,n$.
Here $I_p^n=I_p\times\cdots\times I_p$ and
$J_{p;m}^n=J_{p;m_1}\times\cdots\times J_{p;m_n}$
are the $n$-direct products of the corresponding sets
(\ref{62.0**}) and (\ref{62.0*}).

Using (\ref{62.1-1}), (\ref{9}), it is easy to verify that
\begin{equation}
\label{62.1-2}
\int_{\bQ_p}\Theta_{\gamma s a}^{(m)}(x)\,d^nx=0.
\end{equation}
Thus the functions (\ref{62.8}) are {\it $n$-dimensional $p$-adic
wavelets}. According to (\ref{62.1-2}) and Lemma~\ref{lem1},
$\Theta_{\gamma s a}^{(m)}(x)$ belong to the Lizorkin space
$\in \Phi(\bQ_p^n)$.

For any $\gamma\in \bZ$ and  $s=(s_1,\dots,s_n)\in J_{p;m}^n$
the functions (\ref{62.8}) are periodical with the vector periods
$T_{\gamma s}=(T_{1|\gamma s},\dots,T_{n|\gamma s})\in p^{m-\gamma}\bZ_p^n$.

In view of (\ref{9}), Theorem~\ref{th2} implies the following statement.

\begin{Theorem}
\label{th2.1}
The functions {\rm(\ref{62.8})} form an orthonormal complete
basis in ${\cL}^2(\bQ_p^n)$ {\rm(}$p$-adic wavelet basis{\rm)}.
\end{Theorem}

\begin{Corollary}
\label{cor2}
The functions
$$
{\widetilde\Theta}_{\gamma s a}^{(m)}=F[\Theta_{\gamma s a}^{(m)}](\xi)
=p^{n\gamma/2}\chi_p\big(p^{-\gamma}a\cdot\xi\big)
\Omega\big(|s+p^{-\gamma}\xi|_p\big),\quad \xi\in \bQ_p^n,
$$
form an orthonormal
complete basis in ${\cL}^2(\bQ_p^n)$,
$a=(a_1,\dots,a_n)\in I_p^n$; $s=(s_1,\dots,s_n)\in J_{p;m}^n$;
$m=(m_1,\dots,m_n)$, $m_j\ge 1$ is a {\it fixed} positive
integer, $j=1,2,\dots,n$.
\end{Corollary}

The proof follows from Theorem~\ref{th2.1}, formula (\ref{64.8*})
(see below) and the Parseval formula~\cite[VII,(4.1)]{Vl-V-Z}.

\section{$p$-Adic wavelets as eigenfunctions of pseudo-differential operators}
\label{s6}

\subsection{Pseudo-differential operators.}\label{s6.1}
As mentioned above, the one- dimensional Kozyrev wavelets
(\ref{62.0}) introduced in~\cite{Koz0} is a particular case of
the wavelets (\ref{62.1}) for $m=1$. Moreover, in~\cite{Koz0}
S.~V.~Kozyrev proved that his wavelets (\ref{62.0}) are
eigenfunctions of the one-dimensional Vladimirov operator
$D^{\alpha}$ for $\alpha>0$:
$$
D^{\alpha}\theta_{\gamma j a}(x)=p^{\alpha(1-\gamma)}\theta_{\gamma j a}(x),
\quad x\in \bQ_p,
$$
where $\gamma\in \bZ$, $a\in I_p$, $j=1,2,\dots p-1$. Later, it was
proved in~\cite[4.4.]{Al-Kh-Sh3} that in fact, the Kozyrev wavelets
(\ref{62.0}) are eigenfunctions of the Vladimirov operator for any
$\alpha$, i.e., the above formula holds for all $\alpha\in \bC$.

Now we prove that $n$-dimensional wavelets (\ref{62.8}) are eigenfunctions
for a class of pseudo-differential operators (\ref{64.3}), which includes
the Taibleson fractional operator (\ref{59**}), (\ref{63.4}).

\begin{Theorem}
\label{th4.1}
Let $A$ be a pseudo-differential operator
with a symbol $\cA(\xi)\in \cE(\bQ_p^n\setminus \{0\})$.
Then the $n$-dimensional $p$-adic wavelet {\rm (\ref{62.8})}
is an eigenfunction of $A$ if and only if
\begin{equation}
\label{64.1***}
\cA\big(p^{\gamma}(-s+\eta)\big)=\cA\big(-p^{\gamma}s\big),
\qquad \forall \, \eta \in \bZ_p^n,
\end{equation}
where $\gamma\in \bZ$, $j \in J_{p;m}^n$, $a\in I_p^n$.
Here the eigenvalue $\lambda=\cA\big(-p^{\gamma}s\big)$, i.e.,
$$
A\Theta_{\gamma s a}^{(m)}(x)=\cA(-p^{\gamma}s)\Theta_{\gamma s a}^{(m)}(x).
$$
\end{Theorem}

\begin{proof}
Let $\Theta_{s}^{(m)}(x)=\chi_p(s\cdot x)\Omega(|x|_p)$;
$x\in \bQ_p^n$; $s=(s_1,\dots,s_n)\in J_{p;m}^n$,
$s_k\in J_{p;m_k}$, $k=1,2,\dots,n$.
Using (\ref{9}), (\ref{14.1}), (\ref{14}), we have
$$
F[\Theta_{s}^{(m)}(x)](\xi)
=F\Big[\prod_{k=1}^n\chi_p(x_ks_k)\Omega(|x_k|_p)\Big](\xi)
=\prod_{k=1}^nF\Big[\Omega(|x_k|_p)\Big](\xi_k+s_k|_p)
$$
\begin{equation}
\label{64.7}
=\prod_{k=1}^n\Omega\big(|\xi_k+s_{k}|_p\big)
=\Omega\big(|\xi+s|_p\big), \quad \xi\in \bQ_p^n.
\end{equation}
Here, in view of (\ref{9}), $\Omega\big(|\xi+s|_p\big)
=\Omega\big(|\xi_1+s_{1}|_p\big)\times\cdots\times
\Omega\big(|\xi_n+s_{n}|_p\big)$.

According to (\ref{62.0*}), $|s_{k}|_p=p^{m_k}$, i.e.,
$\Omega\big(|\xi_k+s_{k}|_p\big)\ne 0$ only if
$\xi_k=-s_{k}+\eta_k$, where $\eta_k\in \bZ_p$,
$s_k \in J_{p;m_k}$, \ $k=1,2,\dots,n$. Thus $\xi=-s+\eta$,
where $\eta \in \bZ_p^n$, $s\in J_{p;m}^n$, and in view
of (\ref{8}), $|\xi|_p=p^{\max\{m_1,\dots,m_n\}}$.

In view of formulas (\ref{62.8}), (\ref{64.7}), (\ref{14}), we have
$$
F[\Theta_{\gamma s a}^{(m)}(x)](\xi)
=p^{-n\gamma/2}F[\Theta_{s}^{(m)}(p^{\gamma}x-a)](\xi)
\qquad\qquad\qquad\quad
$$
\begin{equation}
\label{64.8*}
\qquad
=p^{n\gamma/2}\chi_p\big(p^{-\gamma}a\cdot\xi\big)
\Omega\big(|s+p^{-\gamma}\xi|_p\big).
\end{equation}

Let condition (\ref{64.1***}) be satisfied. Then (\ref{64.3}),
(\ref{64.8*}) imply
$$
A\Theta_{\gamma s a}^{(m)}(x)
=F^{-1}\big[\cA(\xi)F[\Theta_{\gamma s a}^{(m)}](\xi)\big](x)
\qquad\qquad\qquad\qquad\qquad\qquad\qquad
$$
\begin{equation}
\label{64.9}
=p^{n\gamma/2}F^{-1}\big[\cA(\xi)\chi_p\big(p^{-\gamma}a\cdot\xi\big)
\Omega\big(|s+p^{-\gamma}\xi|_p\big)\big](x).
\end{equation}
Making the change of variables $\xi=p^{\gamma}(\eta-s)$
and using (\ref{14.1}), we obtain
$$
A\Theta_{\gamma s a}^{(m)}(x)
=p^{-n\gamma/2}
\int\limits_{\bQ_p^n}\chi_p\big(-(p^{\gamma}x-a)\cdot (\eta-s)\big)
\cA(p^{\gamma}(\eta-s))\,\Omega(|\eta|_p)\,d^n\eta
\qquad
$$
$$
\qquad
=p^{-n\gamma/2}\cA(-p^{\gamma}s)
\chi_p\big(s\cdot(p^{\gamma}x-a)\big)
\int_{B_{0}^n}\chi_p(-(p^{\gamma}x-a)\cdot\eta)\,d^n\eta
$$
$$
=\cA(-p^{\gamma}s)\Theta_{\gamma s a}^{(m)}(x).
\qquad\qquad\qquad\qquad\qquad\qquad\qquad\qquad
$$
Consequently,
$A\Theta_{\gamma s a}^{(m)}(x)=\lambda\Theta_{\gamma s a}^{(m)}(x)$,
where $\lambda=\cA(-p^{\gamma}s)$.

Conversely, if
$A\Theta_{\gamma s a}^{(m)}(x)=\lambda\Theta_{\gamma s a}^{(m)}(x)$,
$\lambda\in \bC$, then, using (\ref{64.3}), (\ref{64.8*}), (\ref{64.9}),
we have
$$
\big(\cA(\xi)-\lambda\big)
\Omega\big(|s+p^{-\gamma}\xi|_p\big)=0, \quad \xi\in \bQ_p^n.
$$
The latter equation has a nontrivial solution only if
$s+p^{-\gamma}\xi=\eta$, $\eta \in \bZ_p^n$, i.e.,
$\xi=p^{\gamma}(-s+\eta)$ and
$\lambda=\cA\big(p^{\gamma}(-s+\eta)\big)$ for any $\eta \in
\bZ_p^n$. Thus $\lambda=\cA(-p^{\gamma}s)$, and, consequently,
(\ref{64.1***}) holds.

The proof of the theorem is complete.
\end{proof}

The following particular statement was proved in~\cite{Al-Kh-Sh3}.

\begin{Corollary}
\label{cor3} {\rm (~\cite{Al-Kh-Sh3})} Let $A$ be a {\it
homogeneous\/} pseudo-differential operator with a symbol
$\cA(\xi)\in \cE(\bQ_p^n\setminus \{0\})$ of degree~$\pi_{\beta}$.
Then the $n$-direct product $\Theta_{\gamma j a}(x)$ of the
one-dimensional Kozyrev $p$-adic wavelets {\rm (\ref{62.0})}, i.e.,
the $n$-dimensional $p$-adic wavelet {\rm (\ref{62.8})}
$\Theta_{\gamma s a}^{(1)}$ (for $m=1$) is an eigenfunction of $A$
if and only if
\begin{equation}
\label{64***}
\cA(-p^{-1}j+\eta)=\cA(-p^{-1}j), \qquad \forall \,
\eta \in \bZ_p^n,
\end{equation}
where $\gamma\in \bZ$; $a\in I_p^n$; $j=(j_1,\dots,j_n)$,
$j_k=1,2,\dots,p-1$, $k=1,2,\dots,n$.
Here the eigenvalue
$\lambda=p^{(1-\beta)\gamma}\cA(-p^{-1}j)$, i.e.,
$$
A\Theta_{\gamma j a}(x)=p^{(1-\beta)\gamma}\cA(-p^{-1}j)
\Theta_{\gamma j a}(x).
$$
\end{Corollary}

\subsection{The Taibleson fractional operator.}\label{s6.2}
As mentioned above, the Taibleson fractional operator $D^{\beta}_{x}$
is homogeneous of degree~$\beta$ (see Definition~\ref{de1.1}) and
has a symbol $\cA(\xi)=|\xi|_p^{\beta}$, which satisfies
the condition (\ref{64.1***})
$$
\cA\big(p^{\gamma}(-s+\eta)\big)=|p^{\gamma}(-s+\eta)|_p^{\beta}
=p^{-\beta\gamma}|-s|_p^{\beta}=p^{\beta(\max\{m_1,\dots,m_n\}-\gamma)}
=\cA\big(-p^{\gamma}s\big)
$$
for all $\eta \in \bZ_p^n$. Thus according to Theorem~\ref{th4.1},
the $n$-dimensional $p$-adic wavelet (\ref{62.8}) is an eigenfunction
of $D^{\beta}_{x}$:
\begin{equation}
\label{66}
D^{\beta}_{x}\Theta_{\gamma s a}^{(m)}(x)
=p^{\beta(\max\{m_1,\dots,m_n\}-\gamma)}\Theta_{\gamma s a}^{(m)}(x),
\quad \beta \in\bC, \quad x\in \bQ_p^n,
\end{equation}
$\gamma\in \bZ$, $a\in I_p^n$, $s \in J_{p;m}^n$.

\section*{Acknowledgments}

The authors are greatly indebted to S.~V.~Kozyrev, V.~I.~Polischook,
and M.~A.~Skopina for fruitful discussions.


\begin{thebibliography}{10}

\bibitem{Al-Kh-Sh1} S.~Albeverio, A.Yu.~Khrennikov, V.M.~Shelkovich,
Associated homogeneous $p$-adic distributions,
J. Math. An. Appl. \textbf{313} (2006) 64--83.

\bibitem{Al-Kh-Sh2} S.~Albeverio, A.Yu.~Khrennikov, V.~M.~Shelkovich,
Associated homogeneous $p$-adic generalized functions,
Dokl. Ross. Akad. Nauk \textbf{393} no.~3 (2003), 300--303.
English transl. in Russian Doklady Mathematics.
\textbf{68} no.~3 (2003) 354--357.

\bibitem{Al-Kh-Sh3}
S.~Albeverio, A.Yu.~Khrennikov, V.M.~Shelkovich,
Harmonic analysis in the $p$-adic Lizorkin spaces:
fractional operators, pseudo-differential equations,
$p$-adic wavelets, Tauberian theorems,
Journal of Fourier Analysis and Applications,
Vol. 12, Issue 4,  (2006), 393--425.

\bibitem{Al-Kh-Sh4}
S.~Albeverio, A.Yu.~Khrennikov, V.M.~Shelkovich,
Pseudo-differential operators in the $p$-adic Lizorkin space,
$p$-Adic Mathematical Physics.
2-nd International Conference, Belgrade, Serbia and Montenegro,
15 -- 21 September 2005,
Eds: Branko Dragovich, Zoran Rakic,
Melville, New York, 2006,
AIP Conference Proceedings -- March 29, 2006,
Vol. 826, Issue 1, pp. 195--205.

\bibitem{Al-Kh-Sh5}
S.~Albeverio, A.Yu.~Khrennikov, V.M.~Shelkovich,
$p$-Adic wavelets and linear and semi-linear evolutionary
pseudo-differential equations. (In preparation)

\bibitem{Ar-Dr-V}
I.Ya.~Aref$'$eva, B.G.~Dragovic, and I.V.~Volovich
On the adelic string amplitudes,
Phys. Lett. {\bf B} \textbf{209} no.~4 (1998) 445--450.

\bibitem{Av-Bik-Koz-O}
V.A.~Avetisov, A.H.~Bikulov, S.V.~Kozyrev, and V.A.~Osipov,
$p$-Adic models of ultrametric diffusion constrained by
hierarchical energy landscapes,
J. Phys. A: Math. Gen. \textbf{12} (2002) 177--189.

\bibitem{Ben-Ben}
J.J.~Benedetto, and  R.L.~Benedetto,
A wavelet theory for local fields and related groups,
The Journal of Geometric Analysis \textbf{3} (2004) 423--456.

\bibitem{Ben1}
R.L.~Benedetto,
Examples of wavelets for local fields,
Wavelets, Frames, and operator Theory,
(College Park, MD, 2003), Am. Math. Soc.,
Providence, RI, (2004), 27--47.

\bibitem{Bik-V}
A.H.~Bikulov, and I.V.~Volovich,
$p$-Adic Brownian motion,
Izvestia Akademii Nauk, Seria Math.
\textbf{61} no.~3 (1997) 537--552.

\bibitem{Ge-Sh}
I.M.~Gel$'$fand and G.E.~Shilov,
Generalized functions. vol 1: Properties and operations.
New York, Acad. Press, 1964.

\bibitem{G-Gr-P}
I.M.~Gel$'$fand, M.I.~Graev and I.I.~Piatetskii-Shapiro,
Generalized functions. vol 6: Representation theory and
automorphic functions.
Nauka, Moscow, 1966.

\bibitem{Kh1}
A.~Khrennikov,
p-Adic valued distributions in mathematical physics.
Kluwer Academic Publ., Dordrecht, 1994.

\bibitem{Kh2}
A.~Khrennikov,
Non-archimedean analysis: quantum paradoxes, dynamical systems
and biological models.
Kluwer Academic Publ., Dordrecht, 1997.

\bibitem{Kh4}
A.~Khrennikov,
Information dynamics in cognitive, psychological, social
and anomalous phenomena.
Kluwer Academic Publ., Dordrecht, 2004.

\bibitem{Kh-Koz1}
A.Yu.~Khrennikov, and S.V.~Kozyrev,
Wavelets on ultrametric spaces,
Applied and Computational Harmonic Analysis
\textbf{19}  (2005) 61--76.

\bibitem{Kh-Koz2}
A.Yu.~Khrennikov, and S.V.~Kozyrev,
Pseudodifferential operators on ultrametric spaces and ultrametric
wavelets,
Izvestia Akademii Nauk, Seria Math.
\textbf{69} no.~5 (2005) 133--148.

\bibitem{Kh-Koz3}
A.~Yu.~Khrennikov, and S.~V.~Kozyrev,
Localization in space for free particle in ultrametric quantum mechanics,
Dokl. Ross. Akad. Nauk \textbf{411} no.~3 (2006) 316--322.
English transl. in Russian Doklady Mathematics.
\textbf{74} no.~3 (2006) 906--911.

\bibitem{Koch3}
A.N.~Kochubei,
Pseudo-differential equations and stochastics over
non-archimedean fields,
Marcel Dekker. Inc. New York, Basel, 2001.

\bibitem{Koz0}
S.V.~Kozyrev,
Wavelet analysis as a $p$-adic spectral analysis,
Izvestia Akademii Nauk, Seria Math.
\textbf{66} no.~2 (2002) 149--158.

\bibitem{Koz1}
S.V.~Kozyrev,
$p$-Adic pseudodifferential operators: methods and applications,
Proc.\ Steklov Inst.\ Math.
\textbf{245}, Moscow (2004) 154--165.

\bibitem{Koz2}
S.V.~Kozyrev,
$p$-Adic pseudodifferential operators and $p$-adic wavelets,
Theor.\ Math.\ Physics \textbf{138}, no.~3 (2004) 1--42.

\bibitem{Koz-Os-Av-1}
S.V.~Kozyrev, V.Al.~Osipov, V.C.~A.Avetisov,
Nondegenerate ultrametric diffusion,
J. Math. Phys. \textbf{46} no. 6 (2005) 15 pp.

\bibitem{Liz1}
P.I.~Lizorkin,
Generalized Liouville differentiation and the
functional spaces $L\sb{p}{}\sp{r}(E\sb{n})$. Imbedding theorems,
(Russian) Mat. Sb. (N.S.) \textbf{60}(102) (1963) 325--353.

\bibitem{Liz3}
P.I.~Lizorkin,
Operators connected with fractional differentiation, and classes
of differentiable functions, (Russian)
Studies in the theory of differentiable functions of several
variables and its applications, IV. Trudy Mat. Inst. Steklov.
Vol. 117 (1972), 212--243.

\bibitem{Sam3}
S.G.~Samko,
Hypersingular integrals and their applications.
Taylor \& Francis, London, 2002.

\bibitem{Sam-Kil-Mar}
S.G.~Samko, A.A.~Kilbas, and O.I.~Marichev,
Fractional integrals and derivatives and some of their
applications. Minsk, Nauka i Tekhnika, 1987 (in Russian);
English translation: Fractional integrals and derivatives.
Theory and applications, Gordon and Breach, London, 1993.

\bibitem{Taib1}
M.H.~Taibleson,
Harmonic analysis on $n$-dimensional vector spaces over local
fields. I. Basic results on fractional integration,
Math. Annalen \textbf{176} (1968) 191--207.

\bibitem{Taib3}
M.H.~Taibleson,
Fourier analysis on local fields.
Princeton University Press, Princeton, 1975.

\bibitem{Vl-V-Z}
V.S.~Vladimirov, I.V.~Volovich and E.I.~Zelenov,
$p$-Adic analysis and mathematical physics.
World Scientific, Singapore, 1994.

\bibitem{Vl-V1} V.S.~Vladimirov, I.V.~Volovich,
$p$-Adic quantum mechanics,
Commun. Math. Phys. \textbf{123} (1989) 659--676.

\bibitem{V2}
I.V.~Volovich,
$p$-Adic string,
Class. Quant. Grav. \textbf{4} (1987) L83--L87.

\bibitem{Z1}
W.A.~Zuniga-Galindo,
Pseudo-differential equations connected with $p$-adic forms
and local zeta functions,
Bull. Austral. Math. Soc. \textbf{70} no. 1 (2004) 73--86.

\bibitem{Z2}
W.A.~Zuniga-Galindo,
Fundamental solutions of pseudo-differential operators
over $p$-adic fields,
Rend. Sem. Mat. Univ. Padova \textbf{109} (2003) 241--245.


\end{thebibliography}
\end{document}